# STUDY ON INTERSTITIAL MICROWAVE HYPERTHERMIA WITH MULTI-SLOT COAXIAL ANTENNA

PIOTR GAS

**Key words:** Interstitial microwave hyperthermia, Transverse magnetic (TM) waves, Bioheat equation, Finite element method, Tissue properties, Cancer treatment, Simulation

This study focuses attention on interstitial microwave hyperthermia, which is a popular medical procedure for treatment of pathological human tissues containing tumors. In the paper, the 2D finite element analysis is used to compare the models of the coaxial antenna with one, two and three air slots. The implemented models are based on the wave equation in the TM mode coupled with the Pennes equation under a transient state condition. Moreover, the model takes into account the thermo-electrical properties of human tissue for the antenna operating frequency of 2.45 GHz. Simulation results for different tissues and various multi-slot antenna configurations have been presented. The comparative analysis for single-, double- and triple-slot coaxial antennas shown in this article does not have equivalents in the current literature in the field of hyperthermia.

## 1. INTRODUCTION

In recent years, more and more attention has been devoted to medical applications of electromagnetic fields with regard to both diagnosis and treatment. Healing with electromagnetic radiation has been used for performing thermal therapy, since the electromagnetic waves in the microwave range constitute a source of heat, which can easily penetrate biological structures [1]. Hyperthermia relies on changes in sensitivity to high temperature of healthy and diseased cells. It has been clinically proven that heating in the range 40–45°C can induce apoptosis in tumor cells while not causing any negative changes in normal ones [2]. Higher temperatures lead to the death of all cells, and then we have to deal with the so-called thermal ablation [3]. It is worth mentioning that hyperthermia achieves the positive clinical results and its effectiveness increases in combination with radiotherapy and chemotherapy as well as other specific treatments such as immunotherapy and gene therapy [4]. Different technologies are used to heat up malignant tissues. Development of hyperthermia techniques over the centuries from ancient to modern times seems to be a very interesting subject [5]. Research

AGH University of Science and Technology, Department of Electrical and Power Engineering, al. Mickiewicza 30, 30-059 Krakow, Poland, E-mail: piotr.gas@agh.edu.pl

related to hyperthermia is re-experiencing its renaissance and it is the subject of a growing number of papers [6, 7]. Great hopes for the future involve the utilization of nanotechnology in thermal therapy and the so-called magnetic fluid hyperthermia, where magnetic nanoparticles are injected directly into the place of interest [8]. Concomitant magnetic drug targeting during thermal therapy plays an important role recently as an adjunctive therapy in cancer treatment [9].

Interstitial hyperthermia is a popular medical procedure for heating pathological tissues including tumors and other malignancies located deeply in the human body. Special attention has been focused on this technique, because the small microwave antenna could be directly positioned in the target tissue to provide a suitable gradient temperature in the desired area without overheating its surroundings [10]. High temperature induces thermonecrosis in the cancerous tissues at the distance of 1 to 2 cm around the heat source and therefore this unique technique is suitable for healing tumors less than 5 cm in diameter [9]. Similar minimally invasive interstitial techniques use high frequency needle electrodes, ultrasound transducers, laser fibre optic conductors, or ferromagnetic fluids. The application time of hyperthermia treatment usually varies between 30 – 60 minutes.

This study focuses on interstitial microwave hyperthermia where electromagnetic energy is transmitted to the tissue through the microwave multi-slot coaxial antenna. In the paper, the author presents the numerical models of the coaxial antenna with one, two and three air slots and compares them to each other. The location of the slots in the antenna's structure could be responsible for different energy deposition in the human tissue and thus have a decisive influence on the healing process during the hyperthermia treatment. The simulations have been performed for several tissues such as liver, breast, kidney, lung and brain, which are commonly treated with this method in medical practice [10].

## 2. MODEL DEFINITION

The model geometry of the multi-slot coaxial antenna is presented in Fig. 1. The analysed microwave antenna consists of a central conductor, dielectric, outer conductor and a plastic catheter, which protects the inner elements and serves hygienic purposes. Since the actual antenna is cylindrical and coaxial, 2D axisymmetric model in the cylindrical coordinates ($r$, $\varphi$, $z$) is sufficient for the analysis. It should be pointed out that the computational domain contains only half of the antenna structure and the surrounding human tissue and its external boundary is fixed at $r = 50$ mm. What is more, the antenna is very thin and its width is less than 2 mm. All air gaps are located in the outer conductor at the $z_1 = 36.0$ mm, $z_2 = 40.2$ mm and $z_3 = 44.4$ mm, and have the same size $d_1 = d_2 = d_3 = d$. The distances between the slots are also constant and have a value $l_1 = l_2 = l$.

The antenna dimensions have been derived from the literature [11] and have been gathered together in Table 1.

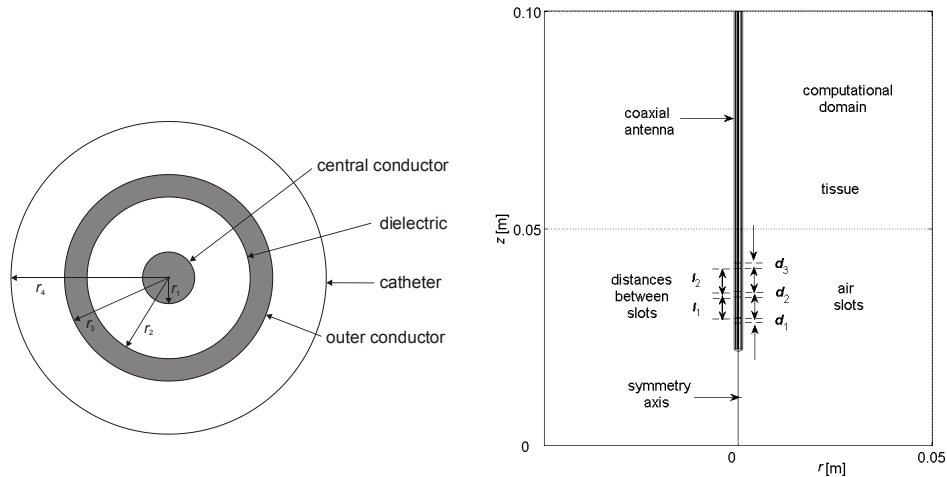

Fig. 1 – Cross section of the multi-slot coaxial antenna (left) and its 2D model with coordinate system and geometrical dimensions (right).

*Table 1*

Dimensions of the microwave antenna in (mm) [11]

| radius of the central conductor | $r_1 = 0.135$ |
|---|---|
| inner radius of the outer conductor | $r_2 = 0.470$ |
| outer radius of the outer conductor | $r_3 = 0.595$ |
| radius of the catheter | $r_4 = 0.895$ |
| size of the air slot | $d = 1.000$ |
| distance between the air slots | $l = 4.200$ |

The interstitial microwave hyperthermia is a complex multiphysics problem, representing the coupling of the electromagnetic and temperature fields. To solve it, two partially independent solutions should be obtained: one for the wave equation and the second for the bioheat equation in transient form. The basic formula for electromagnetic field has been derived from the Maxwell's equations and the material constitutive relations, assuming that the exposed tissues and the individual elements of the multi-slot coaxial antenna are considered as uniform, linear and isotropic media. This fundamental assumption allows for the representation of main quantities of the electromagnetic field using the complex formalism. Considering the above, the wave equation with respect to the vector of the magnetic field intensity **H** takes the following form

$$\text{curl}\,[\,\underline{\varepsilon}_r^{-1}\,\text{curl}\,\boldsymbol{H}\,] - \varepsilon_0\mu_0\mu_r\omega^2\boldsymbol{H} = 0 \tag{1}$$

where $\varepsilon_r$ and $\mu_r$ denote the relative permittivity and relative permeability of the medium, $\varepsilon_0 = 8.854\cdot 10^{-12}$ F/m and $\mu_0 = 4\pi\cdot 10^{-7}$ H/m are the electric and magnetic constants, respectively. Moreover, j denotes the imaginary unit equal to $\sqrt{(-1)}$, $\omega$ is the angular frequency of the electromagnetic radiation and $\underline{\varepsilon}_r$ – the complex relative permittivity defined according the formula

$$\underline{\varepsilon}_r(\omega) = \varepsilon_r - \text{j}\frac{\sigma}{\omega\varepsilon_0} \tag{2}$$

In the above equation, $\sigma$ is the electrical conductivity of the material [S/m].

In the analysed model the electromagnetic waves have the form of transverse magnetic (TM) which means that there are no electromagnetic field changes in the azimuthal direction. Therefore, the magnetic field intensity $\boldsymbol{H}$ has only the $\varphi$-component and the electric field one $\boldsymbol{E}$ propagates in $r$-$z$ plane. This is governed by

$$\boldsymbol{H} = H_\varphi\,\boldsymbol{e}_\varphi\quad,\quad \boldsymbol{E} = E_r\,\boldsymbol{e}_r + E_z\,\boldsymbol{e}_z \tag{3}$$

where $\boldsymbol{e}_r, \boldsymbol{e}_\varphi, \boldsymbol{e}_z$ denote the unit vectors in the cylindrical coordinate system $(r, \varphi, z)$. After taking in into account the above, the wave equation in the described 2D axial-symmetric model takes the scalar form

$$\text{curl}\,[\,\underline{\varepsilon}_r^{-1}\,\text{curl}\,H_\varphi\,] - \varepsilon_0\mu_0\mu_r\omega^2 H_\varphi = 0 \tag{4}$$

Full formulation of the model requires the establishment of the appropriate boundary-initial conditions for the EM and thermal fields. Primarily, on all perfect electrically conducting surfaces the PEC boundary conditions are imposed as

$$\boldsymbol{n}\times\boldsymbol{E} = 0 \tag{5}$$

where $\boldsymbol{n}$ is the unit normal vector perpendicular to the surface. Moreover, at the antenna symmetry axis the boundary conditions in the form of $E_r = 0$ and $\partial E_z/\partial r = 0$ are established. Additionally, the outer boundaries of the computational domain, which represent the tissue boundary, have the so-called scattering boundary conditions. These conditions make the boundaries completely transparent

$$\sqrt{\varepsilon_0\underline{\varepsilon}_r}\,\boldsymbol{n}\times\boldsymbol{E} - \sqrt{\mu_0\mu_r}\,\boldsymbol{H} = -2\sqrt{\mu_0\mu_r}\,H_{\phi0}\,\boldsymbol{e}_\phi \tag{6}$$

where $H_{\varphi 0}$ is the excitation magnetic field on the antenna's input defined by

$$H_{\varphi 0} = \frac{1}{Z\,r}\sqrt{\frac{Z P_{\text{in}}}{\pi\ln(r_1/r_2)}} \tag{7}$$

In the above equation $P_{\text{in}}$ denotes the total microwave input power at the feed point, while $r_1$ and $r_2$ are the dielectric's inner and outer radii, respectively.

Moreover, $Z$ signifies the wave impedance of the dielectric cable according to

$$Z = \sqrt{\frac{\mu_0}{\varepsilon_0 \varepsilon_r}} = \frac{120\pi}{\sqrt{\varepsilon_r}} \qquad (8)$$

The feed point of the microwave antenna is placed at the external boundary of the coaxial dielectric. This point is modeled using a port boundary condition with the total microwave input power set at $P_{in}$ [11, 14].

The problem of heat transfer in biological structures is well described by the Pennes equation [12] that in the transient state is defined by

$$\rho C \frac{\partial T}{\partial t} + \mathrm{div}(-\lambda \,\mathrm{grad}\, T) = \rho_b C_b \omega_b (T_b - T) + Q_{ext} + Q_{met} \qquad (9)$$

where $T$ and $T_b$ denote the tissue and the blood vessel temperatures [K], respectively, $\rho$ and $\rho_b$ – the tissue and the blood densities [kg/m$^3$], $C$ and $C_b$ – the tissue and the blood specific heats [J/(kg K)]. Moreover, $t$ is the time [s], $\lambda$ – the thermal conductivity of the tissue [W/(m K)] and $\omega_b$ – the blood perfusion rate [1/s]. In addition to the important quantities related to the blood flow the model takes into account both the metabolic heat generation rate $Q_{met}$ [W/m$^3$] and the external heat sources $Q_{ext}$ [W/m$^3$] responsible for variations of the temperature inside the tissue, as below

$$Q_{ext} = \sigma \mathbf{E} \cdot \mathbf{E}^* = \sigma |\mathbf{E}|^2 = \sigma E^2 = 0.5\, \sigma\, E_m^2 \qquad (10)$$

where $E$ and $E_m$ denote the effective value and the maximum value of the electric field strength, respectively. The bioheat equation has been solved only in the tissue domain. Since the computational area is truncated to a part of the same tissue, at the outer periphery of the human tissue the adiabatic boundary condition describing thermal insulation is used, according to the following equation

$$\mathbf{n} \cdot (-\lambda\, \mathrm{grad}\, T) = 0 \qquad (11)$$

The initial temperature field is known as the normal temperature of the human body, namely $T = T_0 = 37°C$.

Specification of all physical material properties, the excitation values and other parameters of the model are necessary for simulation. First of all, it was assumed that the antenna operates at the frequency of $f = 2.45$ GHz and the antenna input power is set initially at $P_{in} = 1$ W. The electric properties of the antenna have been borrowed from [11] and have summarised in Table 2. Moreover, Table 3 presents the electro-thermal parameters of the human tissues for the frequency of 2.45 GHz according to [13]. The blood parameters used in the bioheat equation are as follows: $T_b = 310.15$ K, $\rho_b = 1050$ kg/m$^3$, and $C_b = 3617$ [J/(kg K)]. Values of the blood flow rate $\omega_b$ (after unit conversion) and $Q_{met}$ have been taken from [13].

*Table 2*

Electrical parameters of the coaxial-slot antenna [11]

| Antenna elements | $\varepsilon_r$ | $\mu_r$ | $\sigma$ [S/m] |
|---|---|---|---|
| dielectric | 2.03 | 1 | 0 |
| catheter | 2.60 | 1 | 0 |
| air slot | 1 | 1 | 1 |

*Table 3*

Physical parameters of various human tissues [13]

| Tissue | $\varepsilon_r$ | $\sigma$ [S/m] | $\lambda$ [W/(m K)] | $\rho$ [kg/m$^3$] | $C$ [J/(kg K)] |
|---|---|---|---|---|---|
| brain | 44.80 | 2.101 | 0.51 | 1046 | 3630 |
| breast | 57.20 | 1.968 | 0.33 | 1058 | 2960 |
| kidney | 52.74 | 2.430 | 0.53 | 1066 | 3763 |
| liver | 43.03 | 1.686 | 0.52 | 1079 | 3540 |
| lung | 20.48 | 0.804 | 0.39 | 394 | 3886 |

It is worth mentioning that the described model has been adapted from the model for microwave cancer therapy presented in [14]. The basic equations (4) and (9) were solved using the finite element method. For FEM implementation Comsol Multiphysics combined with MATLAB environment have been used. It should be emphasized that the Comsol's model for microwave cancer therapy is the subject of many publications, in which it has been validated with experimental measurements and other numerical methods [11]. However, so far no one has made such as a comparative analysis of the multi-slot coaxial antennas as presented in this paper.

3. **OBTAINED RESULTS**

At the start, the author has investigated whether the thermal equilibrium in the present model is fulfilled and use of the adiabatic boundary condition is justified. The dependences shown in Fig. 2 demonstrate that the temperature in the steady-state at the outer boundaries of the model is slightly above the normal tissue temperature 37°C and its the biggest change does not exceed 0.46 % near the antenna (on the upper edge of the computational area). In the next step, a ceiling levels of the antenna input power for which the temperature in the tissue does not exceed 45°C have been designated. First, a point on the border antenna-tissue with maximum elevation of temperature has been estimated for one slot antenna inserted into the liver tissue and arbitrarily adopted $P_{in}$ = 1 W. Then at this point ($r = r_4 = 0.895$ mm and $z = 37.19$ mm) the temperature distributions depending on the parameter of $P_{in}$ have been determined as shown in Fig. 3. For such specific values of the antenna input power further simulations have been performed.

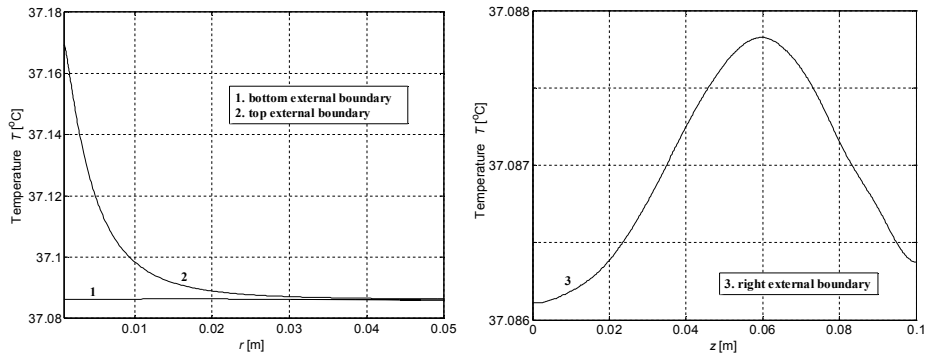

Fig. 2 – Temperature distributions in the steady-state on the external boundaries of the computational domain for paths 1: $z = 0$ m, 2: $z = 0.01$ m (left) and 3: $r = 0.05$ m (right).

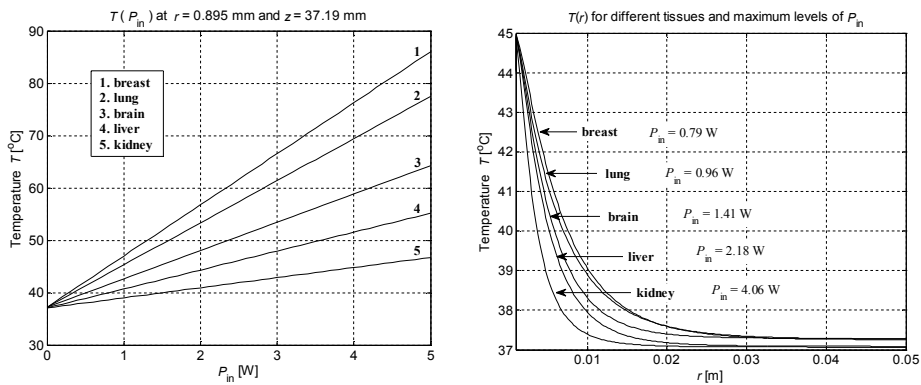

Fig. 3 – Temperature versus the antenna input power at $r = 0.895$ mm and $z = 37.19$ mm (left) and temperature distributions for various human tissues along the path $z = 37.19$ mm (right) in the steady-state for the one slot antenna.

All following simulation results have been calculated for various configurations of active slots and tissues under the same boundary-initial and working antenna conditions (excluding $P_{in}$). Fig. 4 compares together the isotherms within the human breast tissue in the steady-state for single-, double- and triple-slot antennas. A better comparison can be seen in Fig. 5 where the temperature distributions along two well-defined paths are investigated – one passes through the human breast tissue perpendicular to the antenna at the height of the fist air slot ($z = z_1$) and second runs parallel to the symmetry axis at the distance $r = 2.5$ mm.

4222 Piotr Gas 8

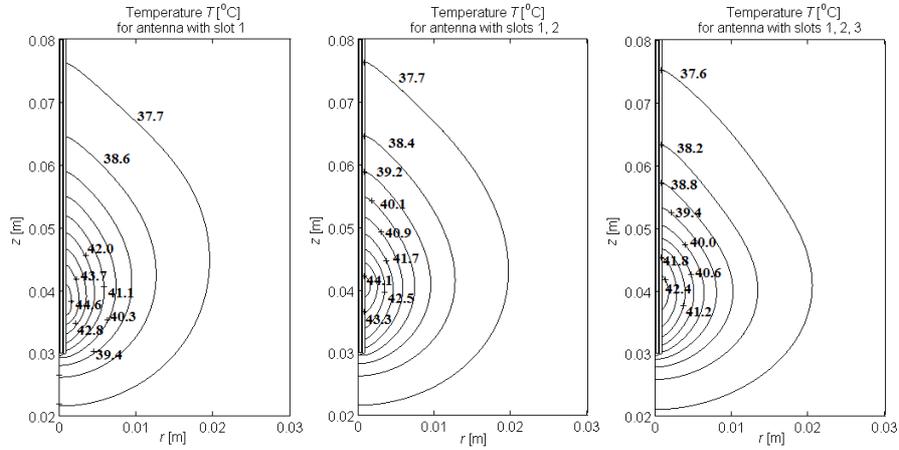

Fig. 4 – Isothermal lines inside the breast tissue in the steady-state for the coaxial antennas with different active slots (all for $f = 2.45$ GHz and $P_{in} = 0.79$ W).

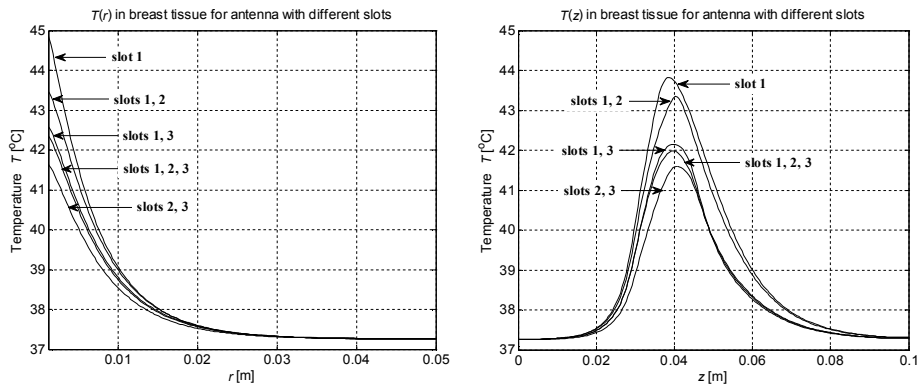

Fig. 5 – Temperature distributions in the human breast tissue along the paths $z = 36$ mm (left) and $r = 2.5$ mm (right) in the steady-state for the coaxial antennas with different active slots.

Time dependent variations of temperature in the case of antennas with different number of active slots inserted into the breast tissue as well as the antenna with slot 1, 2 and 3 inside various human tissues can be seen in Fig. 6. As the point of observation point lying on the height of the second gap ($z = z_2$) at the distance $r = 2.5$ mm from the microwave applicator has been chosen.

Table 4 compares quantitatively all analysed models. $\Delta_r$ denotes the distance from the applicator in which the temperature exceeds 40°C and hyperthermia treatment occurs. Moreover, $z_m$ determines the position of the highest value of temperature on the border antenna-tissue and $T$ is the temperature in the steady-state calculated at point ($r = r_4 = 0.895$ mm and $z = z_m$).

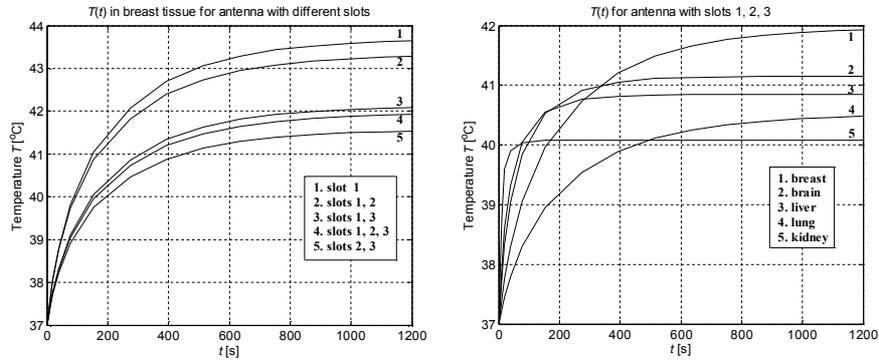

Fig. 6 – Examples of transient temperature distributions at point $r = 2.5$ mm and $z = 40.2$ mm for the coaxial antennas with different active slots in the breast tissue (left) and for the antenna with slots 1, 2, 3 in various tissues (right).

*Table 4*

Quantitative comparison between models ($z_m$, $\Delta_r$ dimensions are in mm and $T$ in °C)

| Tissue | Number of active slot | | | | |
|---|---|---|---|---|---|
| | 1 | 1, 2 | 1, 3 | 2, 3 | 1, 2, 3 |
| **brain** $P_{in} = 1.41$W | $z_m = 37.19$ | $z_m = 40.70$ | $z_m = 43.00$ | $z_m = 43.72$ | $z_m = 39.70$ |
| | $T = 44.93$ | $T = 44.08$ | $T = 42.17$ | $T = 41.67$ | $T = 41.92$ |
| | $\Delta_r = 5.78$ | $\Delta_r = 5.53$ | $\Delta_r = 4.27$ | $\Delta_r = 3.77$ | $\Delta_r = 4.52$ |
| **breast** $P_{in} = 0.79$W | $z_m = 37.69$ | $z_m = 40.70$ | $z_m = 40.00$ | $z_m = 40.20$ | $z_m = 39.70$ |
| | $T = 44.99$ | $T = 44.47$ | $T = 42.78$ | $T = 42.22$ | $T = 42.66$ |
| | $\Delta_r = 7.79$ | $\Delta_r = 7.54$ | $\Delta_r = 6.53$ | $\Delta_r = 5.75$ | $\Delta_r = 6.28$ |
| **kidney** $P_{in} = 4.06$W | $z_m = 36.68$ | $z_m = 40.70$ | $z_m = 43.72$ | $z_m = 40.20$ | $z_m = 39.70$ |
| | $T = 44.99$ | $T = 44.10$ | $T = 41.88$ | $T = 41.30$ | $T = 41.52$ |
| | $\Delta_r = 3.75$ | $\Delta_r = 3.51$ | $\Delta_r = 2.51$ | $\Delta_r = 2.26$ | $\Delta_r = 2.51$ |
| **liver** $P_{in} = 2.18$W | $z_m = 37.19$ | $z_m = 40.70$ | $z_m = 43.72$ | $z_m = 44.22$ | $z_m = 40.00$ |
| | $T = 44.98$ | $T = 44.14$ | $T = 42.16$ | $T = 41.60$ | $T = 44.71$ |
| | $\Delta_r = 5.25$ | $\Delta_r = 5.02$ | $\Delta_r = 1.00$ | $\Delta_r = 3.26$ | $\Delta_r = 3.77$ |
| **lung** $P_{in} = 0.96$W | $z_m = 37.19$ | $z_m = 41.21$ | $z_m = 44.22$ | $z_m = 44.72$ | $z_m = 44.72$ |
| | $T = 44.99$ | $T = 43.68$ | $T = 42.05$ | $T = 41.56$ | $T = 41.69$ |
| | $\Delta_r = 7.03$ | $\Delta_r = 6.53$ | $\Delta_r = 4.52$ | $\Delta_r = 4.02$ | $\Delta_r = 4.27$ |

## 4. FINAL REMARKS

The paper describes the mathematical foundation and presents the numerical simulation results of interstitial microwave hyperthermia with a multi-slot coaxial antenna. The performed analysis showed that adding new slots has no beneficial impact on the therapeutic effect of hyperthermia. As might be expected, the temperature observed in the target area decreases rapidly with the distance from the microwave applicator. The highest temperature occurred within the breast tissue in the case of a single-slot antenna near the air gap. The lowest temperatures were obtained in the kidney tissue for the antenna with slots 2 and 3. Furthermore, the widest ranges of therapeutic area $\Delta_r$ have been obtained for the breast tissue as well as the lung tissue. The temperature stabilized most rapidly in the case of kidney tissue. Of course, in any analysed case, the temperature values can be easily enlarged by raising the level of the antenna's total microwave input power.